\documentclass[aps,prb,reprint,superscriptaddress,showpacs, showkeys, showpacs]{revtex4-2}
\usepackage{hyperref}
\usepackage{graphicx}
\usepackage{amsmath}
\usepackage{amssymb}
\usepackage{color}
\usepackage{upgreek}
\usepackage{bm}
\usepackage{siunitx}
\begin{document}

\preprint{APS/123-QED}

\title{Electron energy-loss spectroscopy on freestanding perforated gold films}

\author{M. Pr\"amassing}
\affiliation{Physikalisches Institut, Rheinische Friedrich-Wilhelms Universit\"at, 53115 Bonn, Germany}
\affiliation{Electron Microscopy and Analytics, Center of Advanced European Studies and Research (caesar), Ludwig-Erhard-Allee 2, 53175 Bonn, Germany.}
\author{T. Kiel}
\affiliation{Institut f\"ur Physik, Humboldt-Universit\"at zu Berlin, Newtonstra\ss{}e 15, 12489 Berlin, Germany.}
\author{S. Irsen}
\affiliation{Electron Microscopy and Analytics, Center of Advanced European Studies and Research (caesar), Ludwig-Erhard-Allee 2, 53175 Bonn, Germany.}
\author{K. Busch}
\affiliation{Institut f\"ur Physik, Humboldt-Universit\"at zu Berlin, Newtonstra\ss{}e 15, 12489 Berlin, Germany.}
\affiliation{Max-Born-Institut, Max-Born-Stra\ss{}e 2A, 12489 Berlin, Germany.}
\author{S. Linden}
 \email{linden@physik.uni-bonn.de}
\affiliation{Physikalisches Institut, Rheinische Friedrich-Wilhelms Universit\"at, 53115 Bonn, Germany}

\date{\today}

\begin{abstract}
We report on a combined far- and near-field study of surface plasmon polaritons on freestanding perforated gold films. The samples are fabricated by focused ion beam milling of {a periodic hole array into} a carbon membrane followed by thermal evaporation of gold and plasma ashing of the carbon film. Optical transmission spectra show a series of characteristic features, which can be attributed to the excitation of surface plasmon modes via the periodic nanohole array. The corresponding near-field distributions are mapped by electron energy-loss spectroscopy.
Besides the optically bright surface plasmon modes, we observe in the near-field an additional dark plasmon mode, which is absent in the normal incidence far-field spectra. Our experimental results are in good agreement with numerical computations based on a discontinuous Galerkin time-domain method.
\end{abstract}

\maketitle


\section{\label{sec:Introduction}Introduction}

Extraordinary optical transmission (EOT) through subwavelength hole arrays in thin metallic films was first demonstrated by Ebbesen et al. in 1998\cite{ebbesen_extraordinary_1998}.
Since then, the fundamental physics leading to this effect has been {investigated} extensively in both theoretical and experimental {studies} \cite{popov_theory_2000,martin-moreno_theory_2001,genet_light_2007,garcia_de_abajo_textitcolloquium_2007,liu_microscopic_2008,braun_how_2009}.
The mechanism behind the EOT phenomenon is the excitation of surface plasmon polaritons (SPPs) upon illumination of the periodic hole array. Due to their surface confinement, the SPPs can reach the other side of the metallic film through the subwavelength holes and couple out via the array again to free space radiation.
The resonant nature of this coupling mechanism makes the EOT phenomenon a narrow-band effect.
Variations of the array geometry parameters open up possibilities to tailor the resonances and utilize the EOT phenomenon for optical color filters \cite{grant_multi-spectral_2016,mccrindle_hybridization_2013} or so called plasmonic printing \cite{cheng_structural_2015}. 
The light confinement provided by the SPPs also makes the structure a good candidate for chemical sensing applications \cite{gordon_new_2008,eftekhari_nanoholes_2009}.
Despite the huge number of fundamental studies regarding theory and experimental far-field measurements, there are only a few reports on experimental near-field studies of comparable structures utilizing scanning near-field optical microscopy \cite{li_scanning_2008,chu_near-field_2007,ctistis_optical_2007,jiang-yan_near-field_2013,mrejen_near-field_2007,sonnichsen_launching_2000} and energy-filtered transmission electron microscopy {(EFTEM)} on oligomer hole structures {\cite{sigle_eftem_2010,talebi_symmetry_2014}}.

Electron energy-loss spectroscopy (EELS) in combination with scanning transmission electron microscopy (STEM) is a powerful near-field characterization technique \cite{von_cube_angular-resolved_2014,von_cube_spatio-spectral_2011,von_cube_isolated_2013,weber_near-field_2017,schroder_real-space_2015,bosman_mapping_2007}. 
STEM-EELS enables near-field mapping of the mode patterns on plasmonic nanostructures over a broad energy range corresponding to wavelengths in the near-infrared and visible  regime. 
In this context, the swift electrons in STEM passing a metallic structure can excite a plasmonic mode.
Thereby the excitation energy corresponds to the kinetic energy-loss of the swift electron. 
The probability of such an event is called electron energy-loss probability (EELP) and is closely related to the electromagnetic local density of states \cite{hohenester_electron-energy-loss_2009,garcia_de_abajo_probing_2008}.
Ultimately, the EELP is related to the induced electric near-field component along the electron beam trajectory.
Raster scanning the electron beam across the sample and recording EEL spectra at different positions allows for a spatio-spectral characterization of the vertical near-field distributions of the plasmonic modes. 

In order to take benefit of EELS as a broadband energy, nanoscale near-field imaging technique requires the samples to be transparent for swift electrons.
To maximize the electron transparency, freestanding perforated metal films are the most desirable choice as a sample system.
Electron transparent perforated metal films have already been fabricated both on SiN-membranes \cite{butun_asymmetric_2015} and completely freestanding.
The freestanding films were fabricated by several different methods like cold-rolling of silver \cite{sigle_eftem_2010,talebi_symmetry_2014}, or transferring the film from a solution onto a supporting grid \cite{cui_directly_2015}. 
Here, we apply a highly reproducible three step method to fabricate freestanding perforated metal films of different thicknesses down to \SI{20}{\nano\meter} \cite{pramassing2019freestanding}.
We characterize the samples both in the far-field with optical transmission spectroscopy and in the near-field with EELS. {The very thin and freestanding nature of the gold films allows us to perform the EELS measurements directly through the film. In contrast, comparable EELS investigations on slit arrays \cite{PhysRevB.93.245417,walther2016coupling} and the EFTEM studies on oligomere hole structures \cite{sigle_eftem_2010} mentioned above were performed with thicker films and could only provide EELS data in the gap regions.}
By comparison of the near- and far-field data we are able to identify a previously undiscovered dark plasmonic mode.
The experimental results are in accordance with numerical computations based on the discontinuous Galerkin time-domain (DGTD) method \cite{busch_dgtd_2011,matyssek_eels_2011}.  

\section{\label{sec:Methods}Methods}

\subsection{\label{subsec:Fabrication}Sample fabrication}

\begin{figure}[ht]
 \centering
\includegraphics[keepaspectratio,width=3.375 in]{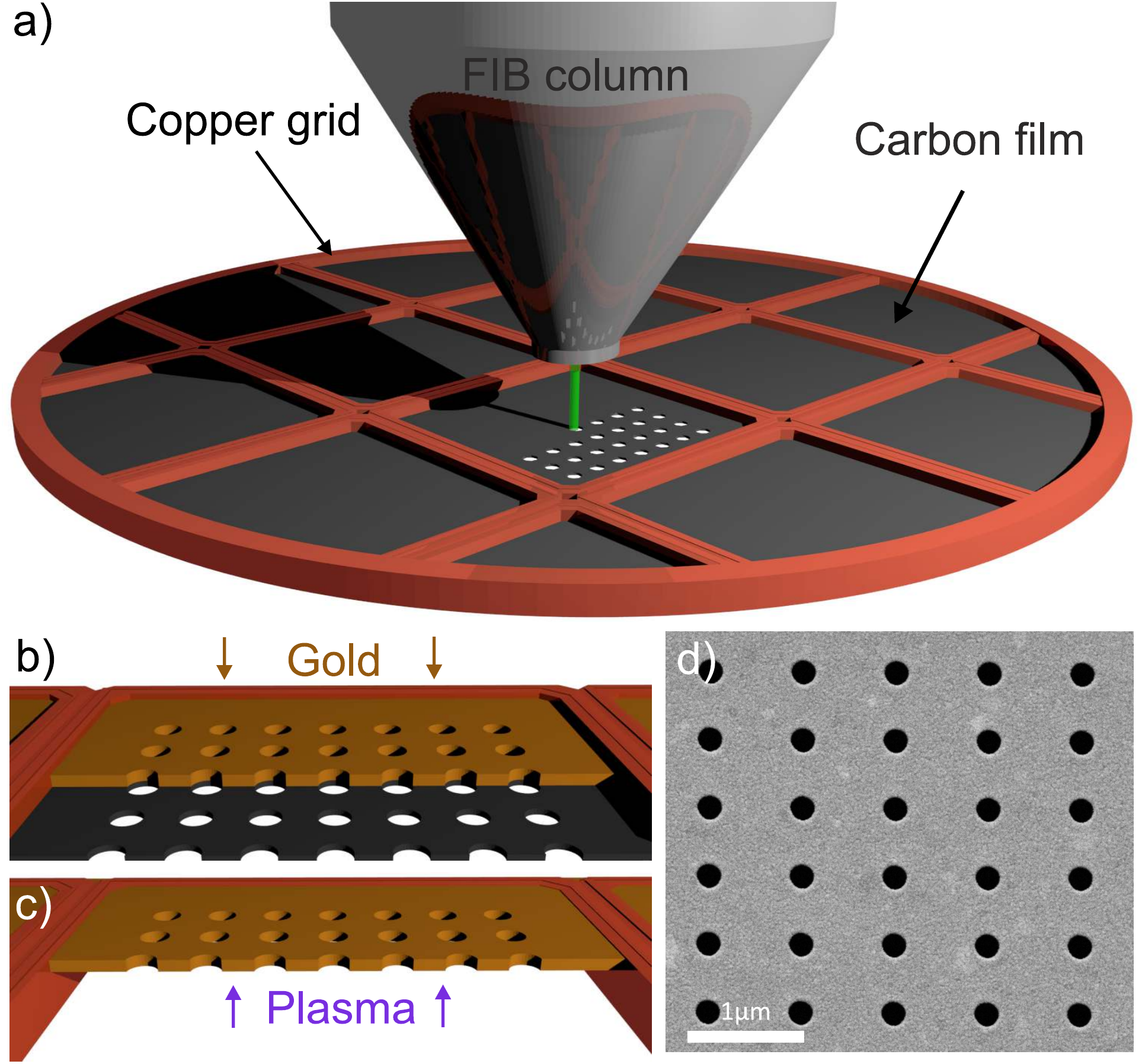}
 \caption{Fabrication process of freestanding perforated gold films: (a) Patterning of a carbon membrane via Focused Ion Beam milling. (b) Thermal evaporation of gold to metallize the structures. (c) Removal of the carbon film by a plasma ashing process. (d) Scanning Electron Microscope viewgraph of the sample after all fabrication steps.}
  \label{fig::Fabrication}
\end{figure}
The sample fabrication process \cite{pramassing2019freestanding} is based on a commercially {available} 
carbon film grid (Quantifoil Cu 200 mesh from Quantifoil Micro Tools GmbH) for application in transmission electron microscopy. 
The approximately \SI{15}{\nano\meter} thick carbon film is patterned by focused ion beam milling with Gallium ions accelerated to \SI{30}{\kilo\electronvolt} with a current of \SI{50}{\pico\ampere} (see Fig.~\ref{fig::Fabrication}). 
Afterwards, the structures are metallized by thermal evaporation of gold with a deposition rate of \SI{2}{\angstrom\per\second}. 
As a last step, the carbon film is removed by a plasma ashing process with a $80:20$ Argon-Oxygen gas compound for \SI{420}{\second} at a pressure of \SI{2.6}{\micro\bar}.
Fig.~\ref{fig::Fabrication}~(d) shows a scanning electron micrograph of a sample after all fabrication steps. 
The samples are fabricated as periodic hole arrays with $p_x = \SI{800}{\nano\meter}$ pitch in horizontal direction and $p_y = \SI{600}{\nano\meter}$ pitch in vertical direction.
The array consists of 100$\times$100 circular holes. 
The hole radius is about R = \SI{100}{\nano\meter} and the film thicknesses range from $d_z= \SI{22}{\nano\meter}$ to $d_z= \SI{74}{\nano\meter}$.

\subsection{\label{subsec:OSA}Optical transmission spectroscopy}

The optical transmission spectra of the perforated films are recorded with a home-build setup.
In this setup, the light of a halogen light bulb is coupled into a multimode optical with a core diameter of \SI{50}{\micro\meter}.
The output of the fiber is collimated, linearly polarized by a Glan Thompson polarizer and imaged onto the sample by a microscope objective $(10\times\text{magnification, NA}=0.1)$. 
The transmitted light is collected by a second microscope objective $(20\times\text{magnification, NA}=0.4)$ and coupled into a grating spectrometer via a multimode fiber.
The spectra are recorded with a Si/InGaAs sandwich diode that gives access to spectral range between \SI{500}{\nano\meter} and \SI{1700}{\nano\meter} wavelength.

\subsection{\label{subsec:EELS}Electron energy-Loss spectroscopy (EELS)}

The STEM-EELS measurements are conducted with a Zeiss Libra200 MC Cs-STEM (CRISP),  which is operated at an acceleration voltage of \SI{200}{\kilo\volt}.
The instrument is equipped with a Cs-corrector for spherical aberration correction of the illumination system.
Furthermore, CRISP includes a second-order aberration corrected $\Omega$-Type monochromator for a well defied initial kinetic energy of the electrons.
After passing the sample, the electron beam is dispersed by a spectrometer onto a $2\text{k} \times 2\text{k}$ SSCCD camera (Gatan, Ultrascan 1000).
The energy resolution of the spectrometer is \SI{16}{\milli\electronvolt}/pixel.
The combination of these techniques guarantees a spatial resolution of a few nanometers combined with an energy resolution of roughly \SI{0.1}{\electronvolt}, which is determined by the FWHM of the zero-loss peak (ZLP).
The latter contains all electrons that have passed the sample without an interaction.
To record EELS maps, the electron beam is raster scanned across the sample with a step size of \SI{4.9}{\nano\meter} per pixel.
A spectrum is recorded in each pixel with an acquisition time of \SI{8}{\milli\second}.
As a postprocessing step, the spectra are separately normalized to their total number of counts and are shifted on the energy axis, such that the ZLP is centered at \SI{0}{\electronvolt}.
Additionally, the background contribution of the ZLP is fitted by a power law and subtracted from the data for each spectrum.

\subsection{Numerical methods}
\label{subsec:Numerics}

\begin{figure}[ht]
 \centering
\includegraphics[keepaspectratio,width=\linewidth]{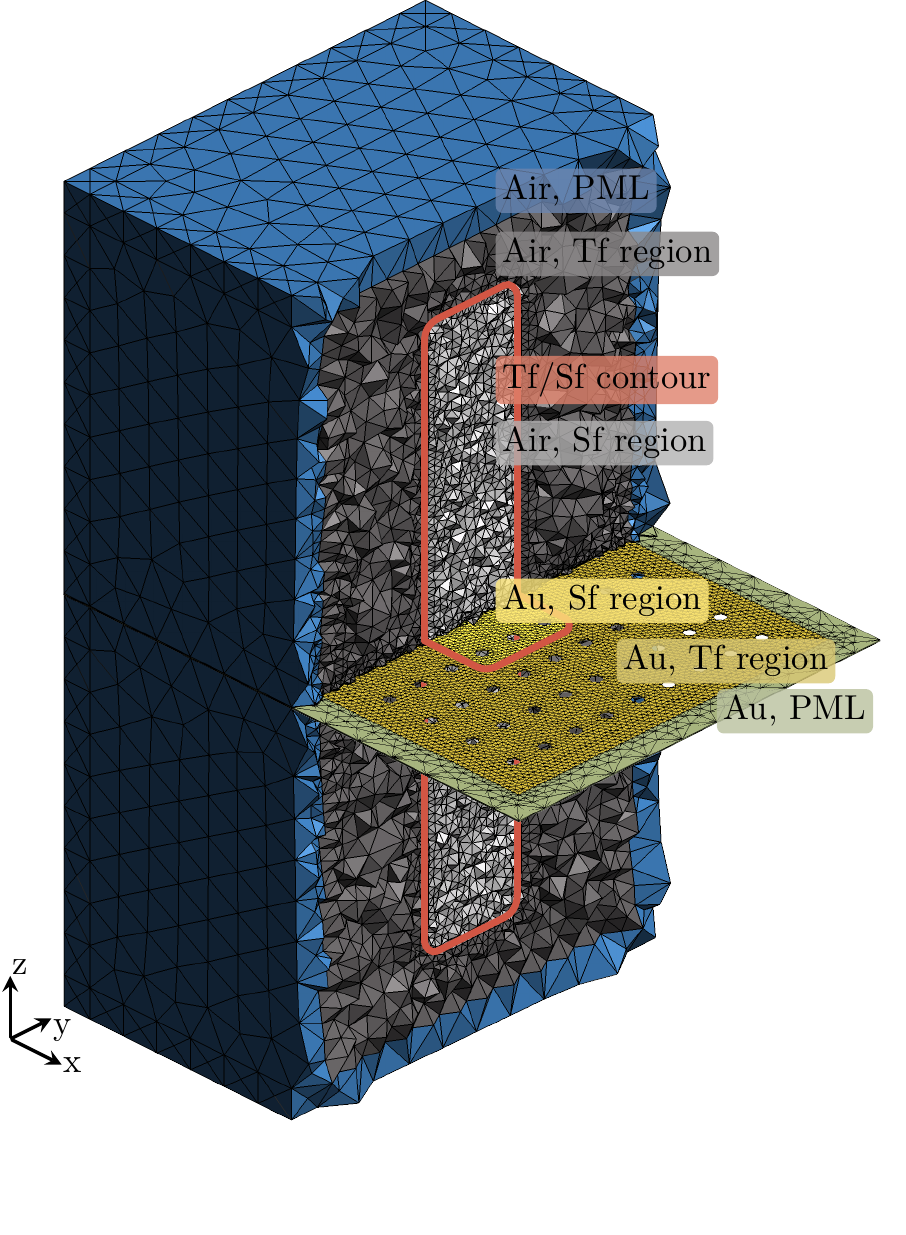}
 \caption{
 Corss section of the finite element mesh used for the DGTD computations.
 This includes a total field (Tf) and scattered field (Sf) regions both for the surrounding air and gold layer.
 The swift electron input field is inject on the Tf/Sf contour and the Au Sf region \cite{busch_dgtd_2011}.
 The computational domain contains a $9 \times 9$ hole array and is terminated with perfectly matched layers (PMLs) to emulate a continuously continued gold film. } 
 \label{fig::EELSmesh}
\end{figure}

The numerical computations of the optical transmittance spectra and the EELP spectra are based on a custom implementation of the DGTD method \cite{busch_dgtd_2011,matyssek_eels_2011}.
For the optical transmittance spectra, we send a pulsed plane wave under normal incidence onto a single $p_x\times p_y = \SI{800}{\nano\meter}\times\SI{600}{\nano\meter}$ unit cell. 
The plane wave is launched using the standard total-field/scatter-field (Tf/Sf) approach \cite{busch_dgtd_2011} from a contour within the half-space above the gold film unit cell. 
The transmittance spectrum is obtained from the transmitted energy flux, which is recorded below the film  and is normalized to the incident flux. 
Thereby, the thickness of the film is varied between $d_z = \SI{22}{\nano\meter}$ to $d_z =  \SI{74}{\nano\meter}$ to match the experimental parameters (see Fig.~\ref{fig::OpticalTransmission}).
While the surrounding air is modelled with $\epsilon = 1$, the gold film is modelled using a Drude-Lorentz model as in Ref. \cite{schroder_real-space_2015}.
In contrast to the plane wave excitation used for the optical transmittance spectra, we use the field of a swift electron moving at $v=0.77\,c$ as source-term for the computations of the EELP spectra. 
The loss probability is computed from the scattered, i.e. the induced field of the electron for the $d_z = \SI{22}{\nano\meter}$ gold film in no-recoil approximation with \cite{ritchie_plasma_1957,garcia_de_abajo_optical_2010}
\begin{align}
  \Gamma (\omega) 
  =
  \frac{e}{\pi \hbar^2 \omega} \int \mathrm{d} t \, \Re \left[ \mathrm{e}^{-\mathrm{i} \omega t} \vec{v} \cdot \vec{E}_{\mathrm{ind}} ( \vec{r}_0 + 
\vec{v} t, \omega) \right] \, .
\end{align}
A variation of the transverse position of the swift electron trajectory yields spatially resolved EELP spectra.
To circumvent the numerical problem of evaluating the diverging electron field when the electron beam hits the material, we impose a radial Gaussian charge distribution with a width of $\sigma = \SI{2}{\nano\meter}$ in the electrons' rest-frame.
Furthermore, the computational domain needs to be extended from a single unit cell using periodic boundary conditions to a super-cell of here $9 \times 9$ unit cells.
Otherwise, every unit cell would be excited by a swift electron simultaneously.
Outside the $9 \times 9$ unit-cell the gold-film is terminated using stretched coordinate perfectly matched layers \cite{chew_3dpml_1994,Koenig_scpmls_2011}.
A cross section of the finite-element mesh discretization used is shown in Fig.~\ref{fig::EELSmesh} and includes elements with side lengths as small as \SI{20}{\nano\meter} to guarantee a sufficient resolution of the holes.
For the expansion of the electromagnetic fields we used a third-order Lagrange polynomial basis.

\section{\label{sec:Results}Results and discussions}

The optical transmission measurements on the hole arrays are performed with a linear polarization of the incident light either in $x$- or in $y$-direction. 
\begin{figure*}[ht]
 \centering
\includegraphics[keepaspectratio,width=5.5 in]{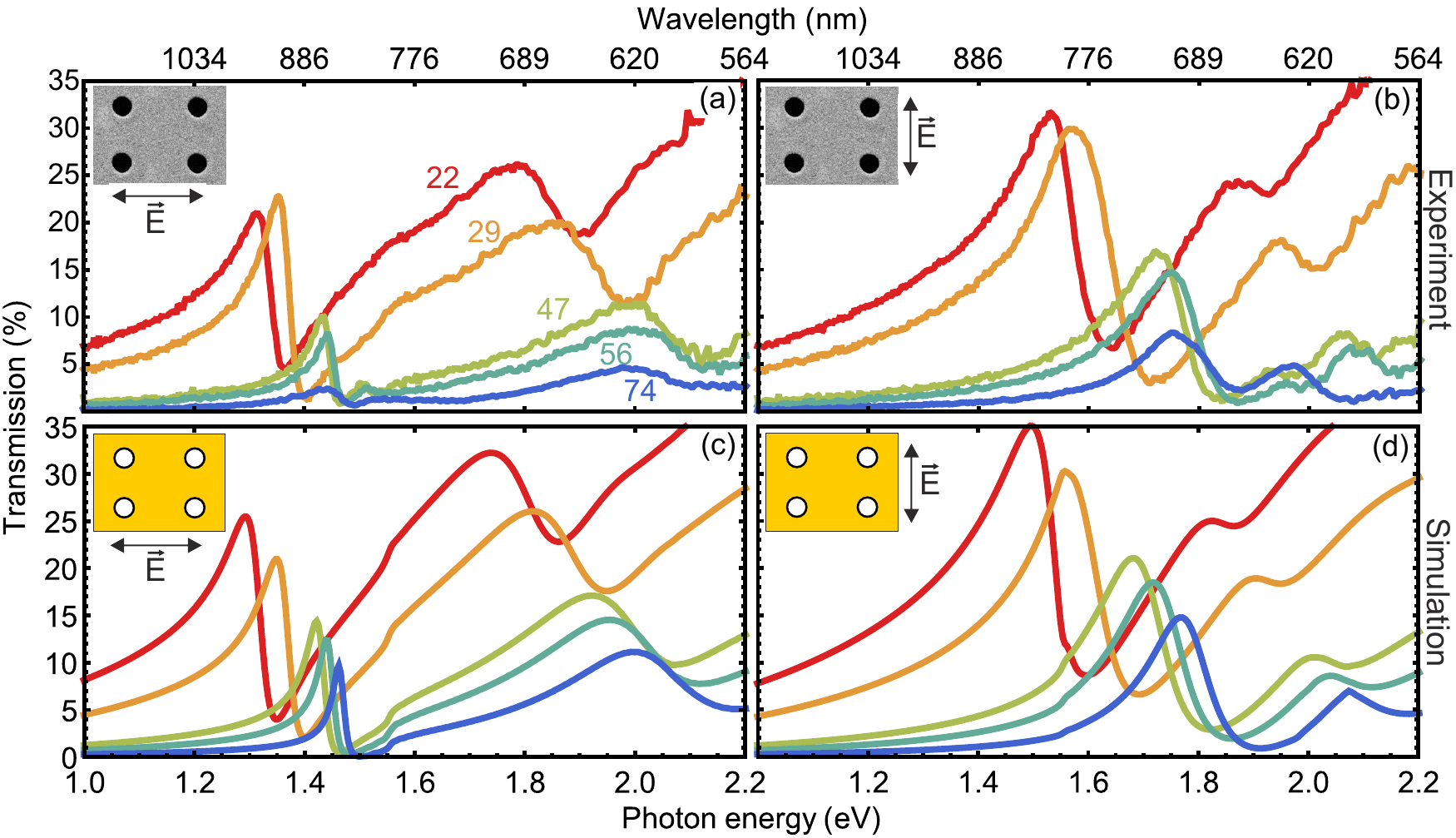}
 \caption{
 Optical transmission spectra of the freestanding hole arrays with different film thicknesses and incident polarizations. The film thicknesses in nanometers are indecated by numbers in the respective color. The electric field is polarized along the horizontal $x$-direction (\SI{800}{\nano\meter} array period) in the left panels and along the vertical $y$-direction (\SI{600}{\nano\meter} array period) in the right panels. Experimental spectra are shown in the upper panels. Computed spectra are shown in the lower panels. 
 }
 \label{fig::OpticalTransmission}
\end{figure*}
Fig.~\ref{fig::OpticalTransmission}~(a) shows the experimental optical transmission spectra of the different films for $x$-polarized excitation.
The film thickness $d_z$ given in nanometers is indicated in each case in the same color as the corresponding transmission spectrum.
Inspecting the red curve for the \SI{22}{\nano\meter} film, one can observe two pronounced resonances around \SI{1.3}{\electronvolt} and \SI{1.8}{\electronvolt}, respectively.
These can be attributed to two different SPP modes resonantly excited by the illumination of the periodic hole array.
For normal incident light, the SPP wavevector has to fulfill the following lattice resonance condition:
\begin{equation}
    \beta=2\pi\sqrt{\frac{m^2}{p_x^2}+\frac{n^2}{p_y^2}},
    \label{eqn::phasematch}
\end{equation}
where $m,n\in \mathbb{N}$. 
Different combinations of $m$ and $n$ yield $(m,n)$-modes at certain resonance energies.
The first observed resonance with the lowest energy at \SI{1.3}{\electronvolt} corresponds to the $(1,0)$ mode of the hole array.
The second resonance around \SI{1.8}{\electronvolt} can be attributed to the $(1,1)$ mode.
The $(1,1)$ mode exhibits a higher overall optical transmission than the $(1,0)$ mode, but the excited resonance is spectrally broader and shows a weaker coupling.
Both observations can be attributed to the fact, that the $(1,1)$ mode at roughly \SI{1.8}{\electronvolt} is closer to the first interband transistion in gold, which naturally leads to a higher direct transmission through the film and higher losses for the excited SPPs.
Fig.~\ref{fig::OpticalTransmission}~(b) shows experimental optical transmission spectra for $y$-polarized incident light.
The first resonance, which lies around \SI{1.55}{\electronvolt} for the \SI{22}{\nano\meter} film corresponds to the $(0,1)$ mode of the array.
As expected, it lies at higher energies than the $(1,0)$ mode due to the smaller period in $y$-direction. 
Additionally, the $(1,1)$ mode can be observed at around \SI{1.9}{\electronvolt}, however with a weaker optical transmission than the $(1,0)$ mode in this case. 
The $(1,1)$ resonance is also weaker compared to the measurements with horizontal polarization and slightly blueshifted.

All resonances show a spectral shift when altering the film thickness.
This can be explained by the SPP dispersion relation for very thin metal films, which accounts for the coupling of the SPPs at the two interfaces.
For a symmetric dielectric environment around the metal film, the dispersion is given by the following implicit relation \cite{yang1991long}:
\begin{equation}
    \tanh(\alpha_m d_z/2) =
    \left(-\frac{\epsilon_d \alpha_m}{\epsilon_m\alpha_d}\right)^{\pm 1}
    \label{eqn::thinfilmspp}
\end{equation}
The thin film supports two propagating SPP modes, namely the short range mode ("$+$" sign in the exponent) and the long range mode ("$-$" sign).
Here, $\epsilon_\text{m}$ and $\epsilon_\text{d}$ denote the dielectric functions of the metal and the surrounding medium, respectively.
Both $\alpha_\text{m}=\sqrt{\beta^2-\epsilon_\text{m} (\omega/c)^2}$ and $\alpha_\text{d}=\sqrt{\beta^2-\epsilon_\text{d} (\omega/c)^2}$ are the decay constants of the SPPs perpendicular to the film.
In the limit of very thin films, the relation requires $\alpha_\text{m} \gg \alpha_\text{d}$ for the long range antisymmetric SPP mode, which effectively leads to a spread of the mode into the dielectric surrounding of the film and a blueshift of the dispersion relation relative to a single interface SPP.
In contrast, the symmetric short range mode requires $\alpha_\text{m} \gg \alpha_\text{d}$, which corresponds to an increased confinement to the metal film and a redshift of the dispersion relation.
Therefore, the enhanced optical transmission is mainly governed by the short range mode via two possible transfer channels:
Firstly, the SPPs feature an enhanced transmission through the holes, enabled by the strong mode confinement as already known for thicker films \cite{genet_light_2007}. Secondly, the short range SPPs enable a transfer through the thin film due to the direct coupling  between SPPs on both interfaces \cite{braun_how_2009}. 
The experimental spectra show the expected redshift of the resonances with decreasing film thickness, due to the energy reduction of the short range SPP dispersion.

In general, the transmission decreases -- as expected -- with increasing film thickness.
An exemption is the \SI{29}{\nano\meter} thick film in the spectral region of the $(1,0)$ mode, where it features a larger transmission than the \SI{22}{\nano\meter} thick film.
This can be attributed to a slightly smaller diameter of the holes in the latter case. 
For the $(0,1)$ and $(1,1)$ mode, there is no higher transmission for the \SI{29}{\nano\meter} film as compared to the \SI{22}{\nano\meter} sample.
The slightly different hole sizes may have smaller impact here, because the SPP wavelength is shorter and the field confinement is higher.
Thus the SPP wave can better pass the smaller holes of the \SI{22}{\nano\meter} sample. 


Fig.~\ref{fig::OpticalTransmission}~(c,d) show the corresponding computed transmission spectra for comparison.
The spectral positions of the two array resonances are in good agreement with the experimental data.
The overall optical transmission in the computed spectra is slightly larger, which indicates a slight underestimation of the losses for the SPPs on the metal film.
The redshift of all resonances with decreasing film thickness is also reproduced by the computations. 

\begin{figure*}[ht]
 \centering
\includegraphics[keepaspectratio,width=6 in]{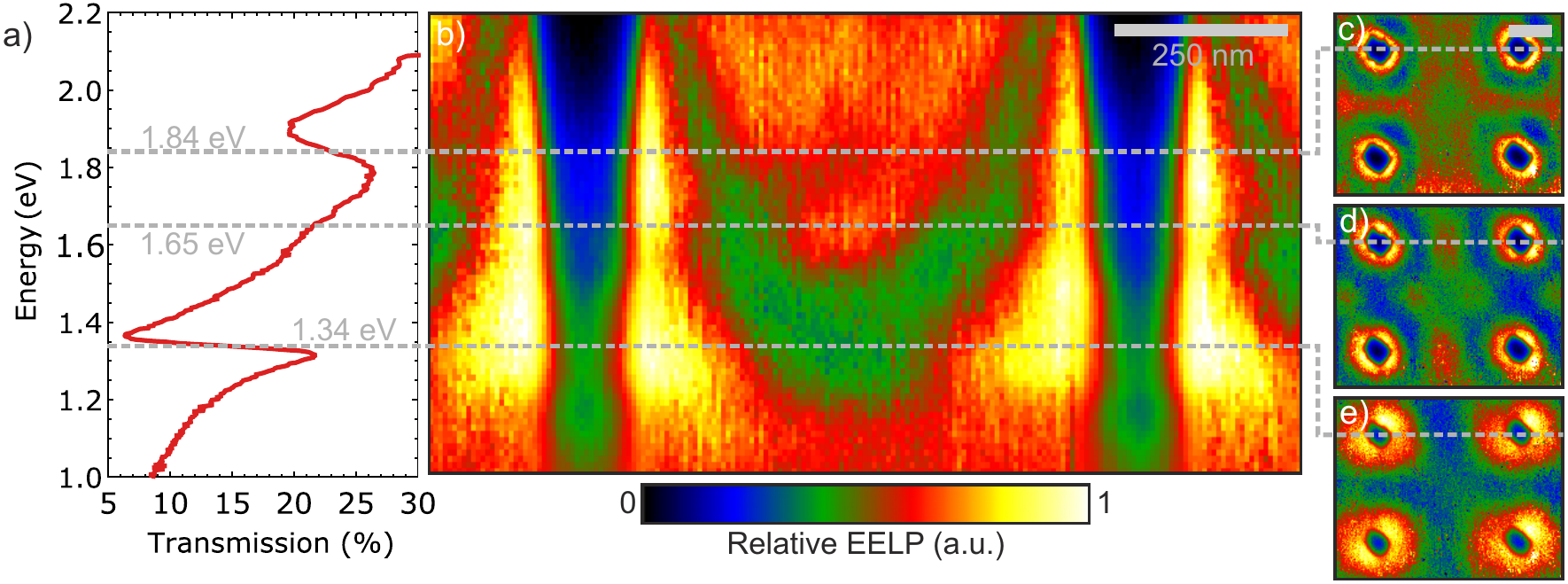}
 \caption{Comparison of an optical transmission spectrum with EELS data for the $d_z =\SI{22}{\nano\meter}$ film. 
 (a) Optical transmission spectrum.
 (b) EELS data extracted from a horizontal line profile through two holes extruded over the energy range of the optical transmission data.
 The grey lines indicate important near- and far-field features.
 (c -- e) 2D EELS maps of the corresponding features in (a,b).
 The scale bar in (c) is \SI{250}{\nano\meter} long.}
 \label{fig::EELS-Data}
\end{figure*}

\begin{figure*}[htb]
 \centering
\includegraphics[keepaspectratio,width=6 in]{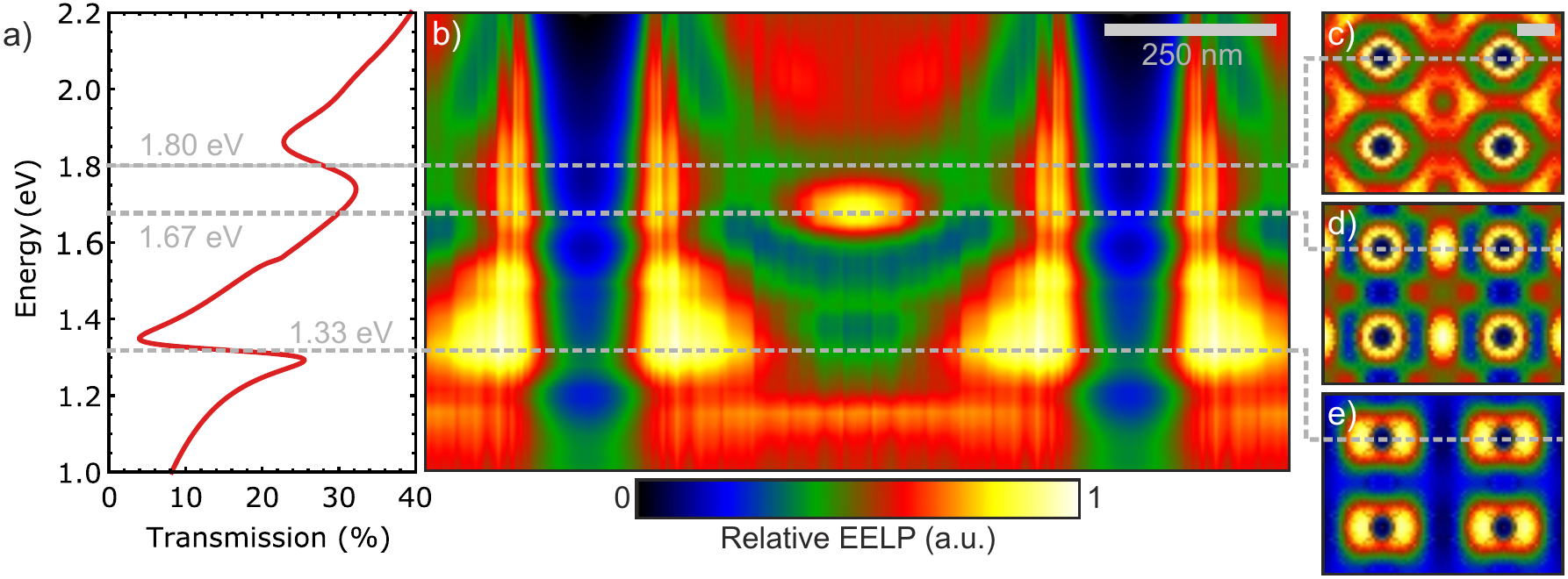}
 \caption{
 DGTD computation results for the optical transmission (a) and EELS data (b -- e) corresponding to the experimental results in Fig.~\ref{fig::EELS-Data}.}
 \label{fig::EELS-Theo}
\end{figure*}

In addition to the optical transmission spectra, we performed an EELS measurements on the thinnest \SI{22}{\nano\meter} film.
The thicker samples are not suited for EELS, since they exhibit a significantly lower transmission for electrons, which hampers a reasonable extraction of the near-field patterns on the film.
Fig.~\ref{fig::EELS-Data}~(a) shows the optical transmission spectrum of the \SI{22}{\nano\meter} film in comparison to the EELP of a line profile through two holes in horizontal direction extruded over the energy range from \SI{1}{\electronvolt} to \SI{2.2}{\electronvolt} ( see Fig.~\ref{fig::EELS-Data}~(b)).
From the data shown in Fig.~\ref{fig::EELS-Data}~(a,b) we can identify three spectral features at the resonance energies of \SI{1.34}{\electronvolt}, \SI{1.65}{\electronvolt} and \SI{1.84}{\electronvolt}, respectively.
Fig.~\ref{fig::EELS-Data}~(c,d,e) show EELP maps of the film in an area surrounding four holes at each energy-loss corresponding to these resonance energies.
Each feature is highlighted by a grey dashed line which indicates its energetic position in (a) and (b), as well as the $y$-position of the profiles in the EELP maps in Fig.~\ref{fig::EELS-Data}~(c,d,e). 

The extracted EELP map for the spectral feature at \SI{1.34}{\electronvolt} in (e) corresponds to the $(1,0)$ mode of the hole array, which is characterized by a standing wave between the holes on the horizontal axis.
The standing wave pattern arises from the excitation of SPP waves at each hole in $\pm x-$direction.
The counterpropagating surface waves interfere on the film and lead to the characteristic near-field pattern, which exhibits two antinodes between a pair of holes, corresponding a phase shift of $2\pi$ between two successive holes as expected for the $(1,0)$ mode. 

The EELS map shows an additional feature at \SI{1.65}{\electronvolt}.
The corresponding near-field pattern in (d) is a standing wave with three antinodes between two holes. 
This corresponds to a phase shift of $3\pi$.
Therefore, this mode can not be excited by a vertically incident plane wave and the spectral feature is absent in the optical transmission spectrum. In contrast, the
swift electron used for the EELS measurement is capable to excite this dark $(1.5,0)$ mode of the array, because it acts as a local white light source for SPPs.

The spectral feature in the optical transmission spectrum occurring at \SI{1.84}{\electronvolt} corresponds to the $(1,1)$ mode of the array.
It has in-plane momentum contributions both in horizontal and vertical directions.
The phase shift between two adjacent holes on the horizontal axis as well as on the vertical axis are expected to be $2\pi$, respectively.
Therefore it can be excited by a plane wave and is observed in the optical transmission spectrum.
The expected phase shift of $4\pi$ on the two diagonal axes between the holes can be observed from the EELS map in Fig.~\ref{fig::EELS-Data}~(c) from the occurrence of four antinodes on these axes.


In order to support the interpretation of the experimental data, regarding the assignment of the modes to the spectral features, Table~\ref{tab:in-plane-momenta} lists the magnitudes of the SPP wavevector required for the different array modes respectively.
\begin{table}[ht]
    \begin{ruledtabular}
    \begin{tabular}{rcc}
        \textbf{Mode} & $\bm{\beta}$ \textbf{(\si{\per\micro\meter})} & $\bm{\lambda_\textbf{SPP}}$ \textbf{(\si{\micro\meter})} \\
        \colrule
        $(1,0)$ & $7.9$ & $0.8$ \\
        $(0,1)$ & $10.5$ &$0.6$ \\
        $(1.5,0)$ & $11.8$ & $0.53$ \\
        $(1,1)$ & $13.1$ &$0.48$\\
        $(0,1.5)$ & $15.7$ &$0.4$\\
        $(2,0)$ & $15.7$ &$0.4$\\
    \end{tabular}
    \end{ruledtabular}
    \caption{
    Resonant SPP wavevector $\beta$ and wavelength $\lambda_\text{SPP}=2\pi/\beta$ calculated by Equation~\eqref{eqn::phasematch} for different array modes $(m,n)$ of the array with $p_x=\SI{800}{\nano\meter}$ and $p_y=\SI{600}{\nano\meter}$.}
    \label{tab:in-plane-momenta}
\end{table}
Since the dispersion relation $E(\beta)$ of the thin film SPPs is monotonically increasing, the order of appearance of the different modes with respect to the resonance energy observed in the experiments should match the order related to the total momentum of the SPP wavevector $\beta$ listed in the table.
Comparing the table to the findings from the experimental investigations, the different modes appear in the expected order.
The modes with in-plane momentum larger than \SI{13.1}{\per\micro\meter} of the $(1,1)$ mode can not be clearly identified from the measurements.
This can be attributed to the fact, that the required mode energy is approaching the first interband transition of gold. 
In order to investigate these modes experimentally one could either enlarge the periods in the array or use silver films. 

To further support the experimental results and confirm the interpretation, we performed numerical computations based on the DGTD method.
The computed EELP spectra and maps are displayed in Fig.~\ref{fig::EELS-Theo} in the same manner as the experimental data.
The spectral features found for the experimental data are reproduced by the computations.
Also the dark $(1.5,0)$ mode is present in the computed EELS data.
All mode energies are in accordance with the experiment within a few tens of \si{\milli\electronvolt}. 

\section{\label{sec:level1} Summary and Conclusions}

\begin{figure}[ht]
 \centering
\includegraphics[keepaspectratio,width=3.1 in]{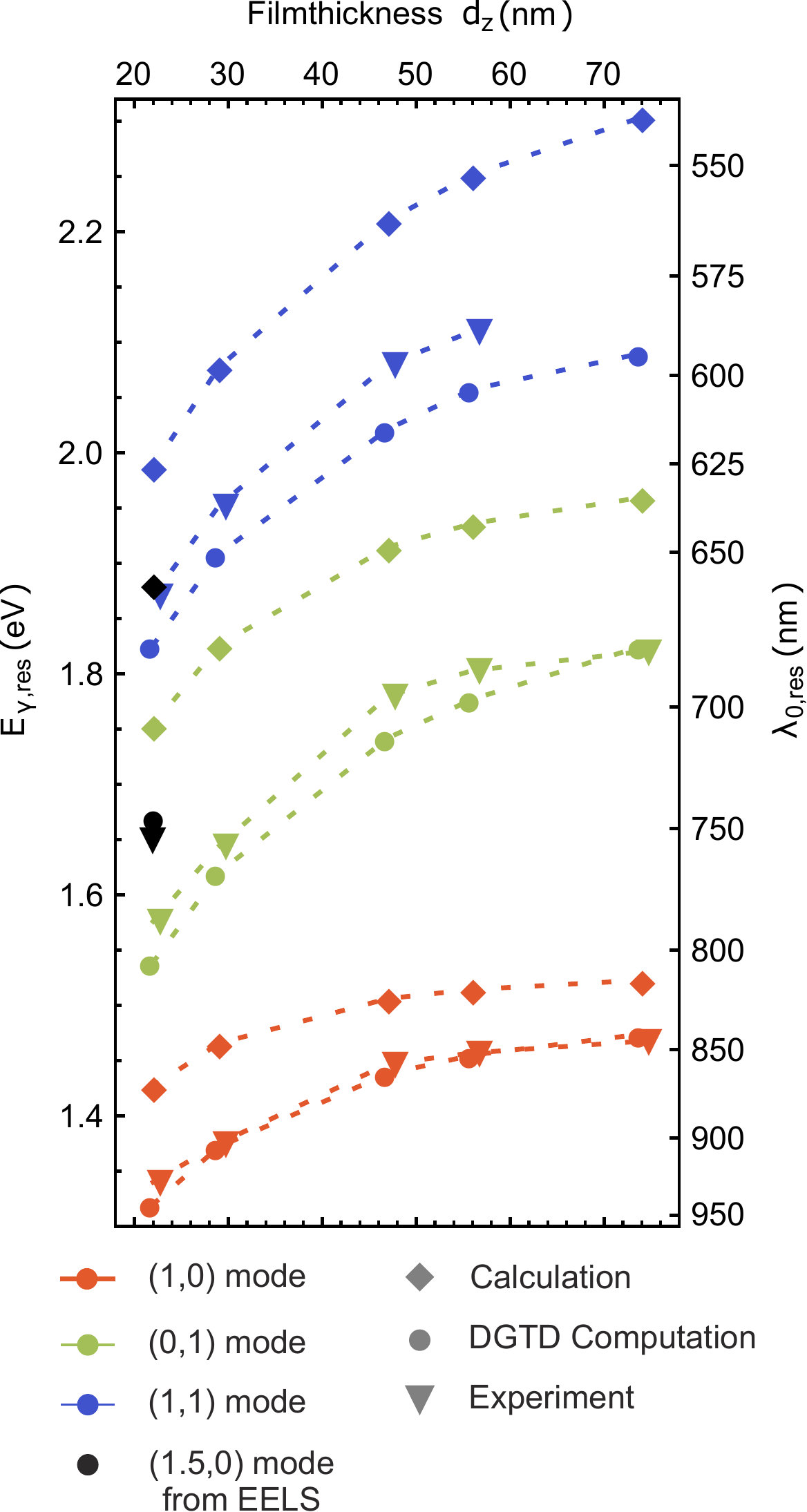}
 \caption{Summarizing graph of the experimental (triangles) and computational (disks) results in comparison with theoretical calculations (rhombs).
 The different colors correspond to the three lowest energy bright modes of the array.
 The black data points represent the dark $(1.5,0)$ mode.}
  \label{fig::EOT_Conclusion}
\end{figure}

To summarize and compare both experimental and computational results, Fig.~\ref{fig::EOT_Conclusion} shows the extracted resonance and loss energies for the $d_z =\SI{22}{\nano\meter}$ thin perforated gold film.
The colored curves show the resonance energies of the three bright modes extracted from all experimental and computed optical transmission spectra.
The resonance positions correspond to the turning point between the minimum and maximum of the corresponding far-field resonance as exemplary given by the gray lines in Fig.~\ref{fig::EELS-Data}~(a) at \SI{1.34}{\electronvolt} and \SI{1.84}{\electronvolt}.
For the $(1,1)$ mode, the extracted values of the measurements and computations with horizontal and vertical incident polarizations are averaged.
The points marked as rhombs are calculated by solving the implicit short range SPP dispersion relation Eq.~\eqref{eqn::thinfilmspp} with the wavevector values $\beta$ from Table~\ref{tab:in-plane-momenta} of the respective array modes.
For all three modes, one can observe a good agreement of the experimental and simulated data.
However, the calculated values exhibit higher resonance energies compared to the measurements and DGTD computations.
These findings can be explained by a shift of the SPP dispersion relation in the presence of the holes with respect to the dispersion relation of a closed film. 
The potential reason is a reduction of the plasma frequency due to the dilution of the metal film by the holes, which leads to a redshift of the SPP dispersion relation.
The three black datapoints correspond to a dark mode of the array, which can only be extracted from the EELS measurements.
This mode can be effectively described by setting $(m,n)=(1.5,0)$ in Equation~\eqref{eqn::phasematch}, corresponding to a phaseshift of $3\pi$ between two adjacent holes on the horizontal axis.
Therefore an optical excitation via a vertical incident plane wave leads to a destructive interference of the SPPs excited by two adjacent holes. 
Hence, the mode can not be excited.
The extracted resonance energies from the experimental and computed EELS data of the thinnest \SI{22}{\nano\meter} film lie between the corresponding energies of the $(0,1)$ and $(1,1)$ mode, as expected from the order in Table~\ref{tab:in-plane-momenta}. 
The same behaviour is also recovered by the calculated dark mode energies.
Like for the bright modes, the calculations predict a higher resonance energy than gained by the experiment and computations. 

In conclusion, plasmonic hole arrays in freestanding thin gold films have been investigated, utilizing EELS as a near-field characterization technique complemented with far-field optical transmission spectroscopy.
The samples feature optical transmission resonances, which occur at the eigenmodes of the plasmonic hole array. 
The standing wave patterns of these eigenmodes could be observed by the near-field measurements and consistently attributed to resonances in the optical far-field transmission spectra. 
Both the EELS and far-field measurements were compared to results from numerical compuations and theoretical calculations. 
From this analysis, we could identify a consistent blueshift of the eigenenergies for all array modes with increasing film thickness in all datasets. 
This behaviour can be explained by a shift of the thin film SPP dispersion relation with decreasing film thickness. 
Interestingly, the EELS analysis revealed an array mode, which could not be observed in the far-field. 
This mode can be classified as an optical dark mode, which can only be excited during the EELS measurement because the electron beam acts as a localized near-field source in this case.

{MP}, SL and SI acknowledge financial support by the Deutsche Forschungsgemeinschaft (LI 1641/5-1).
TK and KB acknowledge
support from the Deutsche Forschungsgemeinschaft (DFG,
German Research Foundation)  - Project-ID 182087777 -
SFB 951.

\bibliography{main}
\end{document}